\documentclass{article} %
\usepackage[final]{colm2026_conference}

\usepackage{microtype}
\usepackage{hyperref}
\usepackage{url}
\usepackage{booktabs}

\usepackage{lineno}

\usepackage{graphicx}
\usepackage{verbatim}
\usepackage{xspace}
\usepackage{amsmath}
\usepackage{listings}
\usepackage{adjustbox}
\usepackage{wrapfig}
\usepackage{tcolorbox}
\usepackage{float}

\usepackage{amsmath,amsfonts,bm}

\def\eqref#1{equation~\ref{#1}}

\def\1{\bm{1}}

\def\vc{{\bm{c}}}

\def\vm{{\bm{m}}}

\def\vx{{\bm{x}}}
\def\vy{{\bm{y}}}

\DeclareMathAlphabet{\mathsfit}{\encodingdefault}{\sfdefault}{m}{sl}
\SetMathAlphabet{\mathsfit}{bold}{\encodingdefault}{\sfdefault}{bx}{n}

\newcommand{\R}{\mathbb{R}}

\title{Routesplain: Towards Faithful and Intervenable Routing for Software-related Tasks}

\author{
    Adam Štorek$^1$\thanks{Work done while interning at Oracle AI. Correspondence to \texttt{astorek@cs.columbia.edu}.}
    \hspace{0.25em}
    Vikas Upadhyay$^2$
    \hspace{0.25em}
    Marianne Menglin Liu$^2$
    \hspace{0.25em}
    Daniel W. Peterson$^2$ \\
    \hspace{0.25em}
    \textbf{Anshul Mittal$^2$}
    \hspace{0.25em}
    \textbf{Sujeeth Bharadwaj$^2$}
    \hspace{0.25em}
    \textbf{Fahad Shah$^2$}
    \hspace{0.25em}
    \textbf{Sujith Ravi$^2$}
    \hspace{0.25em}
    \textbf{Dan Roth$^2$} \\
    $^1$Columbia University
    \hspace{0.25em}
    $^2$Oracle AI \\
}

\begin{document}

\ifcolmsubmission
\linenumbers
\fi

\maketitle

\begin{abstract}
LLMs now tackle a wide range of software-related tasks, yet we show that their performance varies markedly both across and within these tasks. Routing user queries to the appropriate LLMs can therefore help improve response quality while reducing cost. Prior work, however, has focused mainly on general-purpose LLM routing via black-box models. We introduce Routesplain, an interpretable LLM router for software-related tasks, including multilingual code generation and repair, input/output prediction, and computer science QA. Unlike existing routing approaches, Routesplain first extracts human-interpretable concepts from each query (e.g., task, domain, reasoning complexity) and only routes based on these concepts, thereby providing intelligible, faithful rationales. We evaluate Routesplain on 16 state-of-the-art LLMs across eight software-related tasks; Routesplain outperforms individual models both in terms of accuracy and cost, and equals or surpasses all black-box baselines, with concept-level intervention highlighting avenues for further router improvements.
\end{abstract}

\section{Introduction}
Large Language Models (LLMs) are being adopted for an increasing variety of software-related use cases, such as code generation, repair, and test-case generation. As the landscape of available LLMs continues to expand with LLMs offering varying capabilities and pricing tiers, a critical question remains underexplored: do different models perform better or more efficiently on different types of software-related queries?

If such performance variations exist and can be efficiently predicted, they could enable optimization through query-based routing in two scenarios: (a) expert routing for queries requiring specialized capabilities; (b) cost-effective routing when multiple LLMs can adequately answer the query. Prior work has explored routing for general-purpose LLM use cases~\citep{mohammadshahi2024routoo, sikeridis2025pickllm, ong2025routellm}, but has typically focused only on simple code completion scenarios~\citep{hu2024routerbench, zhuang2025embedllm, huang2025lookahead} rather than exploring the rich variability across and within software-related tasks. In this work, we investigate routing efficiencies in accuracy and cost across the diverse landscape of software-related queries, from code generation and repair to execution prediction and computer science question-answering.

Moreover, software queries can be naturally decomposed into \emph{concepts}: high-level, human-understandable characteristics such as which programming languages, libraries, or domain expertise the query requires. We seek to understand whether routing decisions can be made entirely on the basis of such high-level characteristics, without sacrificing performance.
 
We propose Routesplain, an LLM router specifically designed for software-related tasks. Routesplain operates in two independently trained stages: a concept classifier first extracts concepts from the query embedding, and a model classifier then selects the most suitable LLM based solely on these predicted concepts. This design has three direct benefits. First, concepts serve as \textbf{faithful} routing rationales: unlike post-hoc explanations that may not reflect a router's actual reasoning, the concepts that Routesplain surfaces are the same concepts it routes on. Second, routing decisions are \textbf{intervenable}: because concepts are human-readable and editable, users can correct misclassified concepts or explore ``what-if'' scenarios by modifying individual concept values at inference time without retraining. Third, the architecture enables \textbf{diagnostic transparency}: by replacing predicted concepts with ground-truth labels, we can identify which concepts constitute routing performance bottlenecks, an analysis that is impossible with a black-box router.
 
We evaluate Routesplain on a dataset comprising eight software-related tasks across 16 state-of-the-art LLMs including o3 and GPT-4.1. Routesplain outperforms individual models as well as existing black-box routing approaches, achieving accuracy that matches a fully black-box MLP alternative while providing interpretability and control through concept-level interventions. Concept-level intervention reveals that complexity prediction is the primary performance bottleneck (4.05\,pp improvement with oracle labels), while counterfactual experiments confirm that concept changes lead to predictable routing shifts; for example, changing the programming language concept increases the selection probability of top-performing LLMs on the target language by 37\,pp.

Our contributions include: (1) a comprehensive exploration of LLM routing potential within software-related tasks, demonstrating systematic performance variations across programming languages, domains, and natural languages; (2) a routing architecture that decomposes decisions entirely into human-interpretable query concepts without sacrificing performance compared to black-box approaches, providing faithful and intervenable rationales; (3) concept-level intervention and counterfactual validation for LLM routing: new diagnostic methodologies that identify performance bottlenecks and verify routing behavior, revealing that complexity prediction is the primary bottleneck for routing accuracy.

\begin{figure*}
    \centering
    \includegraphics[width=0.8\linewidth]{"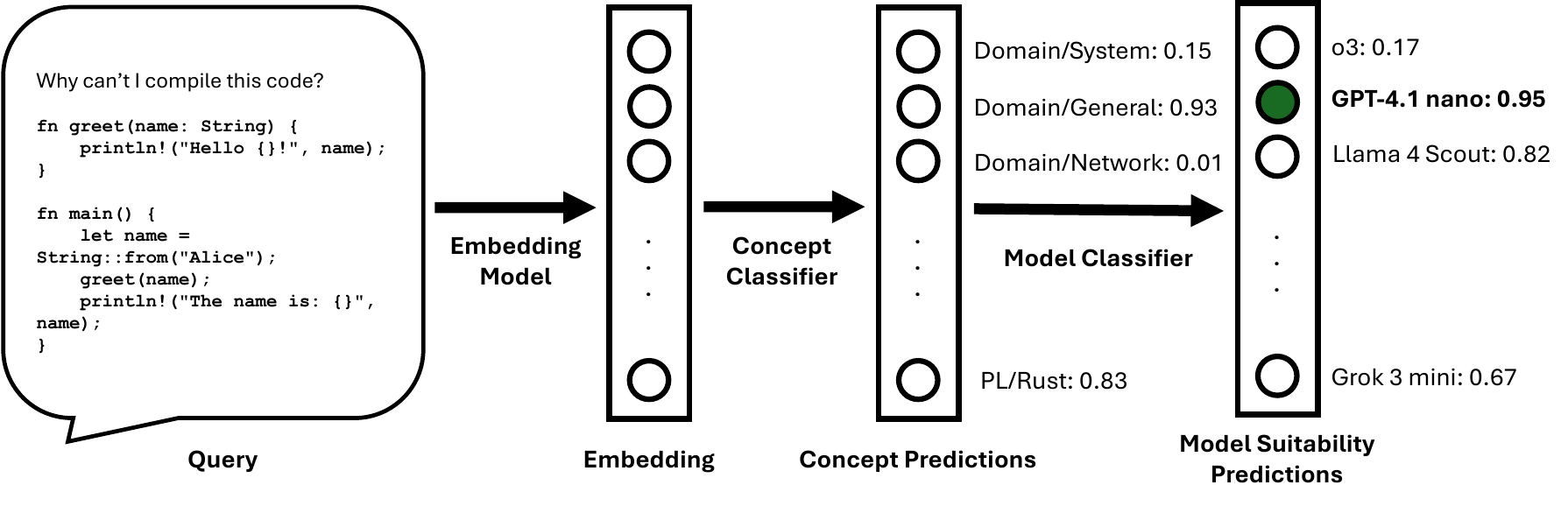"}
    \vspace{-1em}
    \caption{Overview of the Routesplain framework. Each query is embedded using a fixed embedding model. First, a concept classifier takes the contextualized embedding and projects it into the concept space, where each concept represents a high-level, interpretable characteristic of the query (e.g., task, domain, programming language). Then, a separate model classifier takes a concept-space input and outputs model suitability predictions. The query is routed to the most suitable model.}
    \label{fig: routesplain diagram}
    \vspace{-1em}
\end{figure*}

\section{Related Work}
\noindent{\textbf{LLM Ensembling.}} LLM ensemble methods include mixture-of-experts layers~\citep{shazeer2017mixtureofexperts}, speculative decoding~\citep{leviathan2023speculativedecoding}, reranking~\citep{ravaut2022summareranker}, fusing~\citep{jiang2023llmblender}, cascading from weaker to stronger models~\citep{varshney2022cascading, chen2024frugalgpt}, and routing. Routing approaches differ based on (1) learning paradigms, encompassing supervised~\citep{mohammadshahi2024routoo, hu2024routerbench, zhuang2025embedllm}, unsupervised~\citep{guha2024smoothie}, or reinforcement learning~\citep{lu-etal-2024-zooter, sikeridis2025pickllm}; (2) the size of the LLM pool, from a pair of strong and weak LLMs~\citep{ding2024hybridllm, ong2025routellm} to multi-LLM routing~\citep{shnitzer2023routing, sakota2024forc, huang2025lookahead}; (3) routing objectives, with single-objective routing only focused on maximizing accuracy~\citep{zhuang2025embedllm} or multi-objective routing focused on maximizing accuracy while minimizing cost~\citep{mohammadshahi2024routoo, hu2024routerbench}. Concurrent work by~\citet{haque2025codyn} explores LLM routing for coding tasks using Llama 3 models. In contrast, our work demonstrates significant routing potential across model families and generations (Llama 3/4, GPT-4o/4.1, Grok 3/4, o1/o3/o4) and introduces interpretable routing mechanisms for software-related tasks. Particularly relevant to our work, recent approaches focus on implicit~\citep{zhuang2025embedllm} or explicit~\citep{song2025irtrouter} learning of abstract model abilities. Instead, our routing strategy is based on high-level concepts/characteristics that make up the query. \citet{chen2025tagrouter} propose tagging using a generative model and routing based on tag scores. While a compelling direction, a generative approach is both slower than a lightweight classifier and less interpretable due to the large and unconstrained tag space, rendering routing decisions less faithful and harder to correct.

\noindent{\textbf{Model Interpretability.}}
Our goal is to produce routing rationales that are faithful to the router's actual decision process and intervenable at inference time. The closest architectural precedent is the concept bottleneck model~\citep{koh2020conceptbottleneckmodels}, which interposes a human-interpretable concept layer between input and prediction, applied primarily to image classification~\citep{oikarinen2023labelfree, panousis2024coarsetofine}. In that setting, concepts describe input properties to classify \emph{what the input is}. Instead, in our setting, concepts must capture query characteristics that determine \emph{which model should answer the query}. Post-hoc alternatives such as LIME~\citep{ribeiro2016lime}, SHAP~\citep{lundberg2017shap}, model probes~\citep{hewitt2019probes}, and chain-of-thought explanations~\citep{wei2022cot, wang2024cotsft, yue2024mammoth} do not support concept-level intervention. Furthermore, probes can capture representations that are not effectively used by the model~\citep{belinkov2022probing} and CoT-produced rationales have been shown to be unfaithful~\citep{turpin2023unfaithfulcotprompting, chen2024counterfactualsimulatability, chen2025unfaithfulcotreasoning}.

\section{Software-related Task Routing Evaluation \label{sec: dataset}}
\begin{table}
\centering
\begin{adjustbox}{max width=\columnwidth}
\begin{tabular}{l|rrr}
\toprule
       \textbf{Model} & \textbf{Cost (\$/1M)} & \textbf{Avg Len} & \textbf{Total (\$)} \\ \midrule
       o1 & 15.00/60.00 & 147.79 & 557.59 \\
       Grok 3 & 3.00/15.00 & 375.87 & 261.01 \\
       Grok 4 & 3.00/15.00 & 376.66 & 261.47 \\
       GPT-4o & 2.50/10.00 & 266.11 & 138.70 \\ \midrule
       Llama 3.1 405B & 4.00/4.00 & 220.99 & 91.40 \\
       GPT-4.1 & 2.00/8.00 & 188.71 & 87.01 \\
       o3 & 2.00/8.00 & 187.69 & 86.67 \\
       o3 mini & 1.10/4.40 & 208.74 & 51.26 \\
       o4 mini & 1.10/4.40 & 170.63 & 44.78 \\
       GPT-4.1 mini & 0.40/1.60 & 199.58 & 18.07\\ \midrule
       Grok 3 mini & 0.30/0.50 & 294.25 & 9.98 \\
       Llama 4 Mav (fp8) & 0.15/0.60 & 242.39 & 7.77 \\
       GPT-4o mini & 0.15/0.60 & 219.97 & 7.25 \\
       Llama 3.3 70B & 0.13/0.39 & 221.77 & 5.20 \\
       Llama 4 Scout & 0.08/0.30 & 306.20 & 4.70 \\
       GPT-4.1 nano & 0.10/0.40 & 194.82 & 4.44 \\
\bottomrule
\end{tabular}
\end{adjustbox}
\caption{\label{tab: pricing} Model pricing and evaluation costs. Cost shows input/output token pricing per 1M tokens. Average length is measured in tokens per response. Total cost represents the expense of evaluating each model across our dataset collection. Models are grouped by cost-effectiveness tiers: high-cost (top), medium-cost (middle), and low-cost (bottom).}
\vspace{-1em}
\end{table}

In this section, we investigate the extent to which performance variations across LLMs on software tasks justify query-based LLM routing. We examine two scenarios: (1) expert routing, where certain LLMs significantly outperform others on specific queries, and (2) cost-effective routing, where multiple LLMs can answer queries but at different costs. We analyze both \textit{intertask} variability (across different task types like code generation or repair) and \textit{intratask} variability (within tasks based on structural features like programming language or domain), as well as how intratask patterns themselves vary across task types.

\subsection{Routing Dataset Construction \label{subsec: dataset construction}}
Software engineering requires expertise across diverse tasks beyond just code generation, including debugging, code understanding, test synthesis, and domain knowledge. We select tasks that represent these core software engineering capabilities with clear automated evaluation metrics, building upon and expanding the principle of holistic code evaluation~\citep{jain2025livecodebench} that examines multiple dimensions of programming competence.

We assemble popular datasets spanning these capabilities: BigCodeBench~\citep{zhuo2025bigcodebench}, a Python-based, open-domain, function-level code completion and natural-language-instruction-based code generation dataset; CRUXEval~\citep{gu2024cruxeval}, a Python-based, function-level input and output prediction dataset; MMLU-ProX/Computer Science~\citep{xuan2025mmluprox}, the computer science subset of a challenging multiple-choice question dataset translated to 29 natural languages; MultiPL-E~\citep{cassano2023multiple}, a code completion dataset in 24 programming languages.

Recognizing the significance of code repair, following prior work~\citep{olausson2024selfrepair, zheng2025opencodeinterpreter, jain2025livecodebench}, we construct repair tasks by reusing the incorrect responses of permissively licensed LLMs on BigCodeBench. We filter the generated code snippets to include only non-trivially incorrect code, i.e., snippets that execute without crashing but fail to pass the test cases. Furthermore, we deduplicate the code snippets using basic string matching. We formulate two repair tasks: BigCodeBench-Repair, where we provide only the incorrect code snippet, and BigCodeBench-Repair+Info, where we additionally provide execution failure feedback (see~\autoref{appendix: bcb repair templates} for prompt templates).

The entire dataset collection constitutes 38,685 examples, spanning eight verifiable tasks: BCB-Complete (1,140), BCB-Instruct (1,140), BCB-Repair (5,124), BCB-Repair+Info (5,124), CRUXEval-I (800), CRUXEval-O (800), MMLU-ProX/CS (11,890), and MultiPL-E (12,667). We evaluate 16 state-of-the-art LLMs that capture a wide range of capabilities and prices, across multiple families: GPT models (4.1 full/mini/nano, 4o full/mini), Grok models (3 full/mini, 4), o-series models (o1, o3 full/mini, o4 mini), and Llama models (3.1 405B, 3.3 70B, 4 Scout/Maverick fp8). We include a cost breakdown in~\autoref{tab: pricing}. Following prior work~\citep{zhuo2025bigcodebench, roziere2024codellama, liu2023codegen, lai2023ds1000, gu2024cruxeval}, we use greedy decoding and measure pass@1 (2-shot for CRUXEval and 0-shot otherwise).

\subsection{Intertask Performance}
\begin{figure*}
    \centering
    \includegraphics[width=1.0\linewidth]{"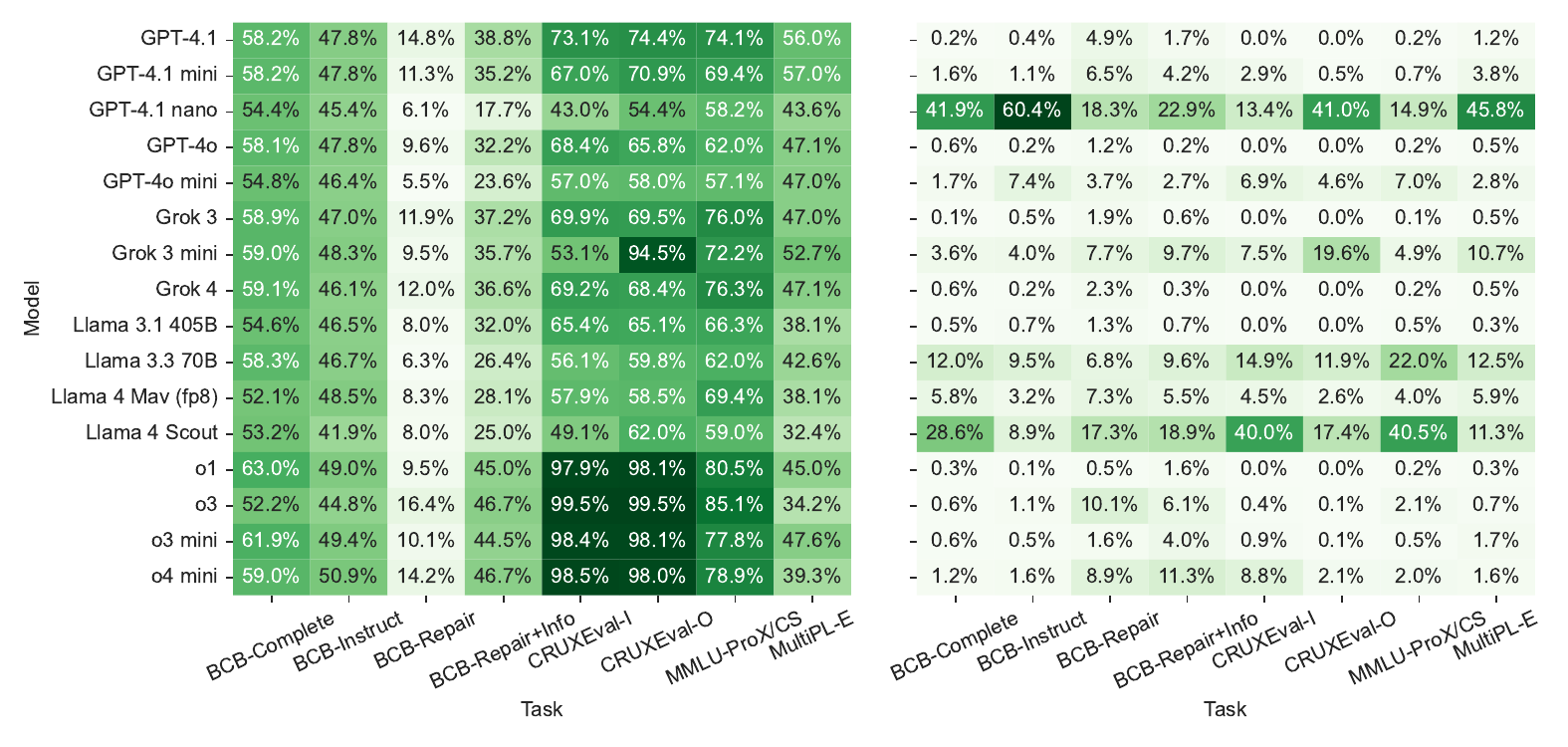"}
    \vspace{-2em}
    \caption{Intertask Performance. \textbf{(a) left:} Pass@1 accuracy of each LLM on each task. \textbf{(b) right:} Share of queries each LLM answered most cost-effectively per task, i.e., correctly at the lowest cost among all correct models.}
    \label{fig: intertask perf}
    \vspace{-1em}
\end{figure*}

\autoref{fig: intertask perf}a shows that o-series reasoning models achieve the highest accuracy on most tasks, with o3 reaching 99.5\% on CRUXEval. The exception is multi-programming-language code completion, where GPT-4.1 mini achieves the highest accuracy at 57.0\%. Nonetheless, \autoref{fig: intertask perf}b reveals that aggregate accuracy obscures significant routing potential. Cost-effective models can handle many queries optimally: GPT-4.1 nano provides the most cost-effective solution for the largest share of queries across BCB-Complete, BCB-Instruct, CRUXEval-O, and MultiPL-E. Simultaneously, expensive models like o3 remain optimal for challenging queries (10.1\% of BCB-Repair), demonstrating the value of expert routing for difficult tasks.

Even within the low-cost tier, different models excel on different tasks. While GPT-4.1 nano dominates most tasks, Llama 4 Scout provides optimal solutions for 40.0\% and 40.5\% of CRUXEval-I and MMLU-ProX/CS queries, respectively, versus only 13.4\% and 14.9\% for GPT-4.1 nano. This suggests that expert routing benefits exist not only between cost tiers, but also within them.

\subsection{Intratask Performance \label{dataset: intratask perf}}
\begin{figure*}
    \centering
    \includegraphics[width=1.0\linewidth]{"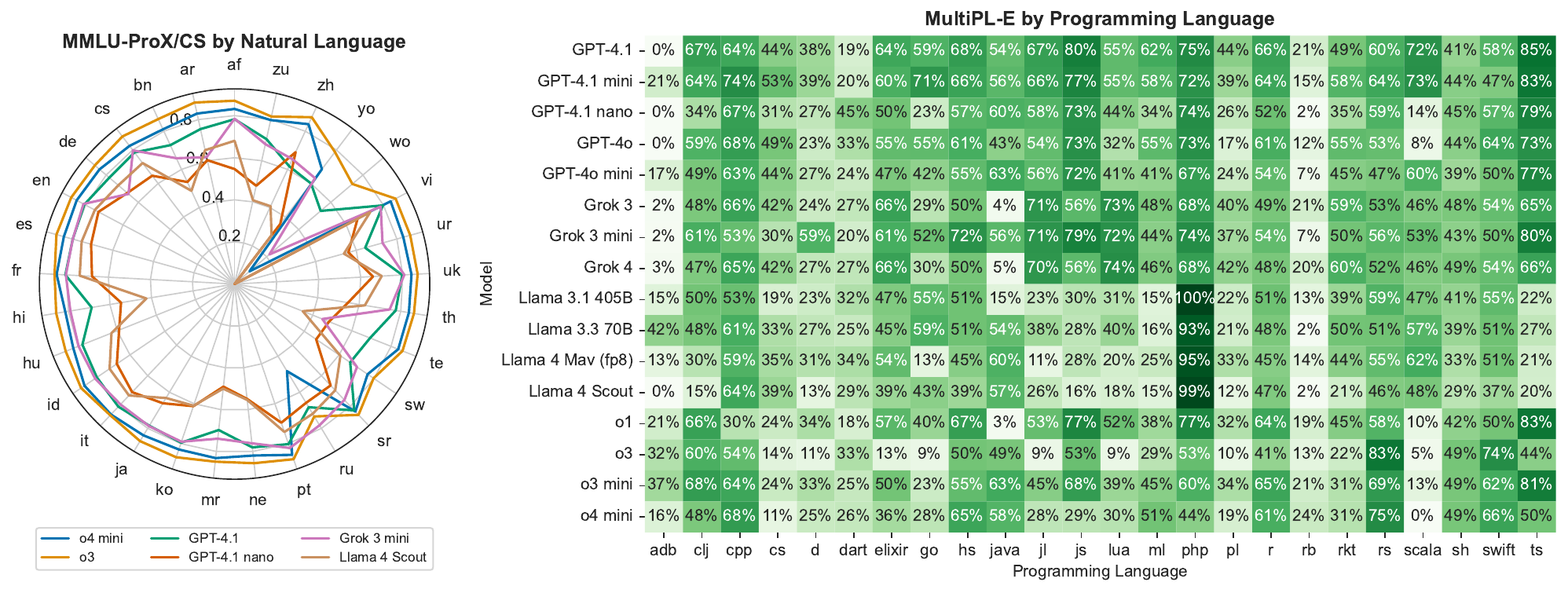"}
    \vspace{-2em}
    \caption{Pass@1 intratask accuracy. \textbf{(a) left:} Computer-science QA, stratified by the natural language of the query, for six representative models. \textbf{(b) right:} Multi-programming-language code completion, stratified by the programming language of the query.}
    \label{fig: intratask perf nl pl}
    \vspace{-1em}
\end{figure*}

\noindent{\textbf{Natural and Programming Language Variations.}} \autoref{fig: intratask perf nl pl}a shows that LLM accuracy varies dramatically based on natural language. While o3 performs consistently well across most languages, it performs worse on Russian (similarly to other OpenAI models GPT-4.1 and o4-mini) and is outperformed by Grok 3 mini. Lower-cost models match the accuracy of higher-cost models like o3 and GPT-4.1, especially on Western European languages like English, Spanish, French, or Italian. The performance gap between lower- and higher-cost models widens for South Asian languages like Hindi, Marathi, or Telugu. The accuracy of all but the pricier models like o3 and GPT-4.1 drops significantly for Wolof, a low-resource language.\footnote{We adopt the geographic and resource classification of natural languages from~\citet{xuan2025mmluprox}.}

In contrast to natural language patterns, programming language performance reveals more complex specialization patterns (\autoref{fig: intratask perf nl pl}b). Each model family demonstrates distinct advantages: Llama models excel at PHP (all achieving above 93\% accuracy), GPT-4 models perform strongly in TypeScript, Grok models have advantages in Julia and Lua, and o-series models show strength in Rust and Swift. Surprisingly, mini versions sometimes dramatically outperform their full-size counterparts: Grok 3 mini achieves 52 percentage points higher accuracy than Grok 3 on Java, and GPT-4o mini similarly outperforms GPT-4o on Scala.

\noindent{\textbf{Domain-Specific Performance.}}
\begin{figure*}
    \centering
    \includegraphics[width=1.0\linewidth]{"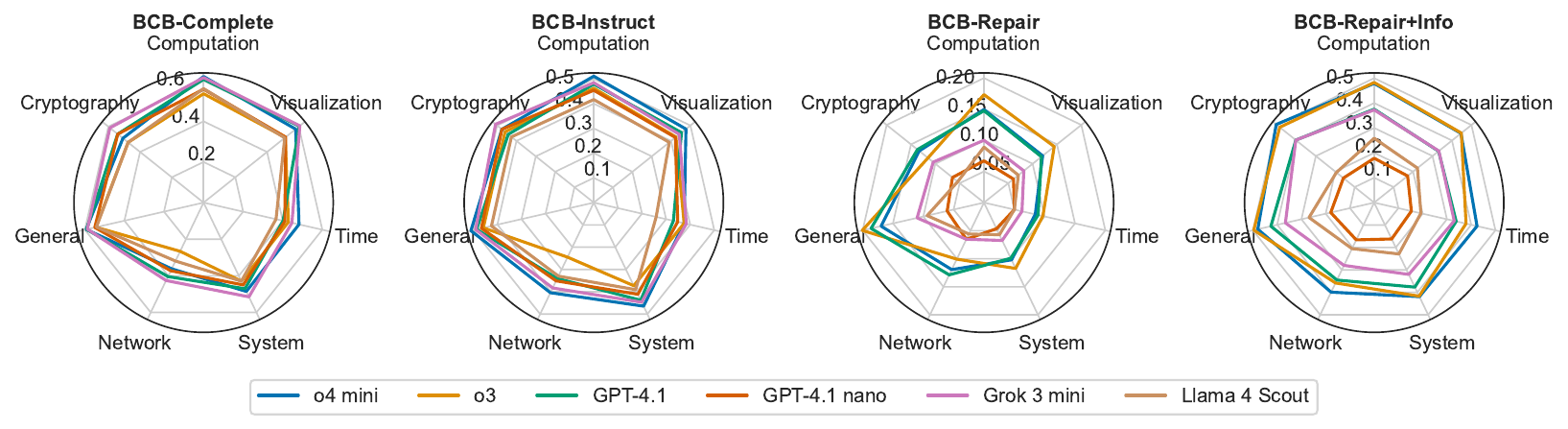"}
    \vspace{-2em}
    \caption{0-shot pass@1 intratask accuracy comparison across open-domain code generation and repair tasks, stratified by the domain of the query, for six representative models.}
    \label{fig: intratask perf domain}
    \vspace{-1.2em}
\end{figure*}
\autoref{fig: intratask perf domain} reveals nuanced domain effects that vary across task types. Consistent with the intertask patterns in~\autoref{fig: intertask perf}, LLMs show much more similar performance to each other on original BigCodeBench tasks (code completion and instruction-based generation) compared to the derived repair tasks, where model performance becomes more stratified. Across all task types, models perform particularly well on generation- and computation-related domains but struggle more with network- and time-related tasks. The repair tasks, however, exhibit notable differences depending on whether execution feedback is provided. Code repair with execution information reduces domain variability across all LLMs, while repair without execution feedback exacerbates these differences. Without execution feedback, models show stronger performance on general- and computation-related snippets but struggle with cryptography- and system-related code, despite not being disproportionately weak at generating such code initially. This pattern reveals interesting model-specific reversals: both o3 and Llama 4 Scout underperform on cryptography- and network-related repair tasks, where they are superseded by their cheaper alternatives o4 mini and GPT-4.1 nano, respectively.

Our analysis demonstrates that LLM performance varies systematically not only across tasks, but also within tasks along high-level query characteristics such as programming language, natural language, and domain. These characteristics are dimensions along which LLMs systematically differ in both accuracy and cost-effectiveness, making them natural candidates for concept-based routing, which we formalize in~\autoref{sec: router formulation}.

\section{Router Formulation \label{sec: router formulation}}
The LLM routing problem can be formulated as learning a function $f: \mathcal{X} \rightarrow \mathbb{R}^n$ that maps an input string $x \in \mathcal{X}$ to a vector $\vm \in \mathbb{R}^n$, where $\vm_i$ corresponds to the probability that LLM $i \in \{1,...,n\}$ provides the most suitable response to $x$. The input $x$ is first projected into a $d$-dimensional contextualized embedding space by an embedding model, so the function to learn becomes $f: \mathbb{R}^d \rightarrow \mathbb{R}^n$, where $\vx \in \mathbb{R}^d$.

A standard supervised approach trains a multi-label classifier that maps from input $\vx \in \mathbb{R}^d$ to model suitability scores $\vm \in \mathbb{R}^n$ given true model correctness labels $\vy \in \{0,1\}^n$, where $\vy_i = 1$ if model $i$ correctly answers the query and $0$ otherwise. The classifier minimizes binary cross-entropy loss with cost regularization (\autoref{eq: opaque router cls}). The cost term represents expected routing cost under the predicted probability distribution, where $cost(\vx^i)$ is a normalized cost vector for input $\vx^i$, and $\lambda$ controls the router's cost sensitivity.
\begin{equation}
\hat{f} = \arg \min_f \sum_i (L_{BCE}(f(\vx^i), \vy^i) + \lambda f(\vx^i)cost(\vx^i))
\label{eq: opaque router cls}
\end{equation}
Our concept-based approach decomposes $f$ into two networks: $f(\vx)=g\circ h(\vx)$.
Motivated by the performance variation patterns observed in~\autoref{sec: dataset}, we leverage existing benchmark metadata such as programming language tags in MultiPL-E and domain and library annotations in BigCodeBench to create a $k$-dimensional concept vector $\vc$ capturing natural language, programming language, domains, required libraries, and three complexity measures. The complexity measures represent the fraction of models that failed on each input, stratified by model type: reasoning complexity (fraction of reasoning models that failed), general complexity (fraction of non-reasoning models that failed), and total complexity (fraction of all models that failed). Reasoning and non-reasoning models form mutually exclusive categories within our 16-model evaluation set.

Network $h$ maps embeddings to concepts: $\mathbb{R}^d \rightarrow \R^k$, trained via~\autoref{eq: concept cls}. Network $g$ maps concepts to model suitability $\mathbb{R}^k \rightarrow \mathbb{R}^n$, trained with cost-regularization via~\autoref{eq: router cls}. We train $g$ and $h$ independently rather than end-to-end. During inference, $\hat{g}$ receives predicted concepts from $\hat{h}$, enabling human intervention by editing concept predictions. We discuss the impact of concept-level intervention in~\autoref{eval: correction}.

\noindent
\begin{minipage}{0.40\linewidth}
\begin{equation}
\hat{h} = \arg \min_h \sum_i L_{BCE}(h(\vx^i), \vc^i)
\label{eq: concept cls}
\end{equation}
\end{minipage}%
\hfill
\begin{minipage}{0.60\linewidth}
\begin{equation}
\hat{g} = \arg \min_g \sum_i (L_{BCE}(g(\vc^i), \vy^i) + \lambda g(\vc^i)cost(\vx^i))
\label{eq: router cls}
\end{equation}
\end{minipage}

\section{Evaluation}
\begin{figure*}
    \centering
    \includegraphics[width=1.0\linewidth]{"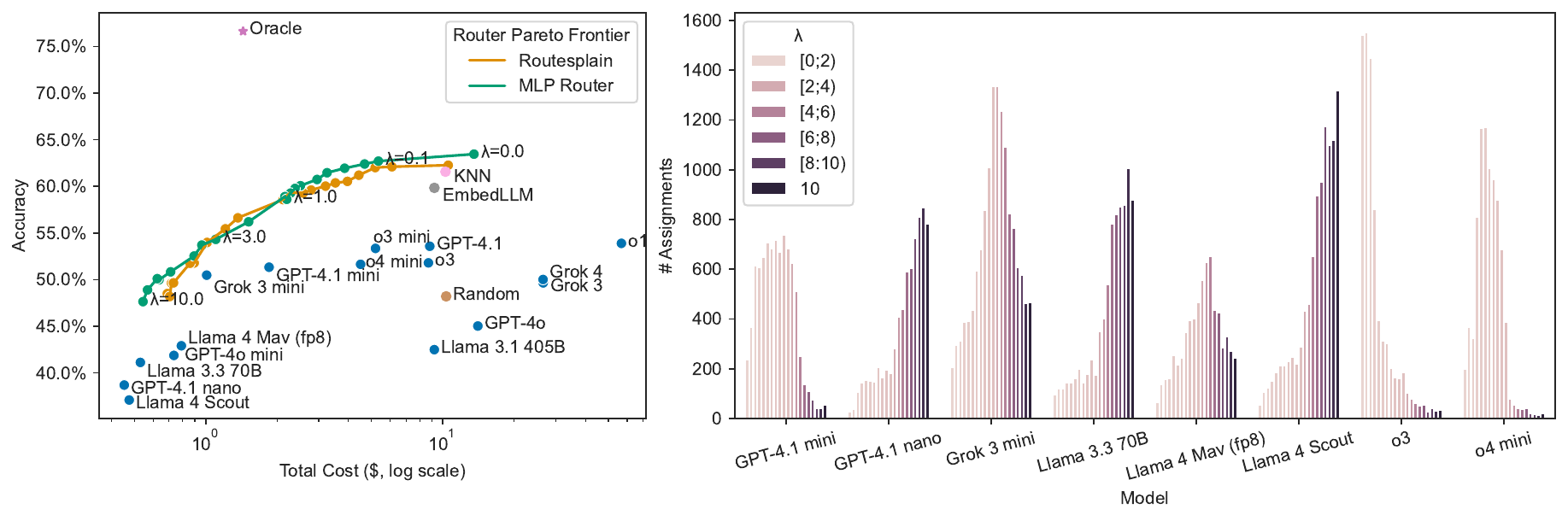"}
    \vspace{-2em}
    \caption{\textbf{(a) left:} Accuracy against cost of each individual model as well as all routers on the test set. For EmbedLLM, MLP router, and Routesplain, we show accuracy and cost averaged across 5 training runs. For Routesplain and MLP router, we display the Pareto frontier established by connecting the average router performance across multiple values of the cost regularizer $\lambda$. \textbf{(b) right:} The average number of times each model was assigned to an input by Routesplain, as we increase the $\lambda$ regularizer. We increase $\lambda$ by $0.1$ on the interval $[0; 1)$ and then by $1$ on the interval $[1;10]$. We show the top 8 most assigned models.}
    \label{fig: frontier}
    \vspace{-1em}
\end{figure*}

We evaluate the efficacy of Routesplain using an 80/10/10 train/validation/test split on our dataset collection (see~\autoref{subsec: dataset construction}). As the collection spans multiple programming and natural languages, we concatenate embeddings from two models: codesage-small-v2~\citep{zhang2024code} and multilingual-e5-small~\citep{wang2024multilinguale5textembeddings}. To control for embedding quality, all baselines use the same embeddings. We retrain each model five times and report mean and standard deviation. Implementation and hyperparameter search details are provided in~\autoref{appendix: impl details}. All experiments are conducted on a single NVIDIA A100 80GB GPU.

Routesplain adds negligible overhead to the routing pipeline. The router itself (462K parameters) processes the entire test set (3,869 queries) in under 10 milliseconds; end-to-end throughput is dominated by embedding time, averaging 104.68 queries per second over 10 runs with unoptimized batching, surpassing most proposed routing solutions by~\citet{ong2025routellm}. Since Routesplain takes only one GPU-minute on a single A100 to retrain from scratch, the retraining cost when the model pool changes is minimal.

\subsection{How does Routesplain compare to individual models and black-box SOTA routers? \label{eval: baseline comparison}}
We compare Routesplain against 16 individual models and three black-box baselines: MLP router, KNN router (both standard baseline types proposed by~\citet{hu2024routerbench}), and EmbedLLM~\citep{zhuang2025embedllm}. For the MLP router, we use the black-box formulation in~\autoref{eq: opaque router cls}. For the KNN router, we train separate binary classifiers for each model to predict correctness, routing to the model with the highest predicted success probability. For EmbedLLM, we reimplement the algorithm as presented in~\citet{zhuang2025embedllm}. All methods receive equal hyperparameter tuning effort. We also include random and oracle routers to establish theoretical bounds. To evaluate performance as cost sensitivity ($\lambda$) changes, we retrain Routesplain and MLP router with $\lambda$ incremented by 0.1 from 0 to 1, and by 1 from 1 to 10. This step size adjustment reflects the greater differences in cost-effectiveness among medium- and low-cost LLMs.

\noindent{\textbf{Routesplain consistently outperforms all 16 individual models}}, achieving higher accuracy at comparable or lower cost across the cost spectrum (\autoref{fig: frontier}a). At 39.34\% lower cost than GPT-4.1, Routesplain ($\lambda=0.1$) achieves an 8.52\,pp higher accuracy. Even at the low-cost tier, Routesplain ($\lambda=4.0$) improves over Grok 3 mini by 3.50\,pp at comparable cost. Routesplain ($\lambda=4.0$) also achieves higher accuracy than the best-performing individual model o1, while being more than $56\times$ cheaper. \autoref{fig: frontier}b shows how Routesplain navigates the cost-accuracy tradeoffs: as cost regularization increases, assignments shift from expensive models (o3) toward mid-tier options (GPT-4.1 mini, o4 mini) and finally to the most cost-efficient models (GPT-4.1 nano, Llama 4 Scout), consistent with the performance patterns identified in~\autoref{sec: dataset}.

\noindent{\textbf{Routesplain Pareto dominates EmbedLLM, KNN router, and random router}}
(\autoref{fig: frontier}a). Notably, Routesplain matches the performance of the MLP router --- the strongest black-box baseline --- at most regularization levels; the MLP router achieves statistically significantly higher mean accuracy only at $\lambda$ values of 0.0, 0.3, and 0.4, while Routesplain statistically significantly outperforms the MLP router at $\lambda=3.0$ (see~\autoref{app: routesplain mlp acc comp}). Routesplain achieves this parity while using 15.54\% fewer parameters (462K vs 547K) and providing the interpretability and intervention capabilities described in the following subsections.

\subsection{How does each concept category contribute to Routesplain's accuracy?}
\begin{table*}
\centering
\begin{adjustbox}{max width=\textwidth}
\begin{tabular}{c|rrr|rrr}
\toprule
& \multicolumn{3}{c|}{\textbf{Concept Group Ablation Acc (\%)}} & \multicolumn{3}{c}{\textbf{Concept Group Intervention Acc (\%)}} \\ \midrule
       \textbf{Concept Group} & $\lambda = 0.0$ & $\lambda = 0.1$ & $\lambda = 4.0$
       & $\lambda = 0.0$ & $\lambda = 0.1$ & $\lambda = 4.0$
       \\ \midrule
       \textit{Baseline}     & 62.27 \tiny (0.32) & 62.10 \tiny (0.55) & 53.98 \tiny (0.39) &
       62.27 \tiny (0.32) & 62.10 \tiny (0.55) & 53.98 \tiny (0.39) \\
       Tasks                 & 62.27 \tiny (0.32) & 62.02 \tiny (0.22) & 53.48 \tiny (0.41) &
       62.30 \tiny (0.30) & 62.25 \tiny (0.63) & 53.89 \tiny (0.37) \\
       Domains               & 62.19 \tiny (0.32) & 62.37 \tiny (0.28) & 53.81 \tiny (0.32) &
       62.27 \tiny (0.33) & 62.27 \tiny (0.68) & 53.92 \tiny (0.42) \\
       Libraries             & 61.69 \tiny (0.28) & 61.55 \tiny (0.32) & 52.56 \tiny (1.37) &
       62.26 \tiny (0.32) & 62.30 \tiny (0.61) & 53.92 \tiny (0.39) \\
       Natural Languages     & 62.11 \tiny (0.16) & 62.15 \tiny (0.15) & 52.71 \tiny (0.60) &
       62.26 \tiny (0.34) & 62.25 \tiny (0.65) & 53.90 \tiny (0.41) \\
       Programming Languages & 60.43 \tiny (0.23) & 60.76 \tiny (0.42) & 51.08 \tiny (0.86) &
       62.27 \tiny (0.33) & 62.13 \tiny (0.67) & 53.98 \tiny (0.41) \\
       Complexity            & 61.45 \tiny (0.29) & 61.28 \tiny (0.19) & 48.43 \tiny (0.69) &
       66.32 \tiny (0.15) & 65.89 \tiny (0.17) & 57.02 \tiny (0.19) \\
\bottomrule
\end{tabular}
\end{adjustbox}
\vspace{-0.5em}
\caption{\label{tab: ablation intervention} Concept group ablation and intervention results across cost regularization levels. \textbf{Ablation:} Accuracy when removing each concept group during training (compared to baseline). \textbf{Intervention:} Accuracy when replacing predicted concept labels with gold labels at inference. Values show mean routing accuracy (\%) with standard deviation over 5 runs.}
\vspace{-1em}
\end{table*}

To understand which concepts contribute most to routing decisions, we conduct ablation studies by removing each concept group and retraining Routesplain across three cost regularization levels ($\lambda = 0, 0.1, 4.0$). Results are shown in~\autoref{tab: ablation intervention} (left column).

\textbf{Concept importance varies significantly with cost sensitivity.} At high cost sensitivity ($\lambda = 4.0$), complexity concepts show the largest impact (-5.55 pp), aligning with the critical need to distinguish expensive versus cheap model requirements when cost matters most. Programming language distinctions matter substantially at all cost levels, reflecting the distinct performance profiles across models shown in~\autoref{dataset: intratask perf}, but have the greatest impact at the lowest cost level. Natural language distinctions show insignificant impact at higher cost levels but meaningful effects (-1.27 pp) when cost is prioritized. This pattern reflects that expensive models handle most natural languages competently, while cheaper models struggle with low-resource languages (\autoref{dataset: intratask perf}). Library concepts are most important at high cost sensitivity and show greater significance than domain concepts, likely because they provide more granular, specific information than broader domain categories.

\textbf{Task categories show minimal ablation effects}, which we attribute to limited cross-dataset label variability in our current collection, as most tasks lack multiple degrees of freedom across natural and programming languages, and domains. Developing richer benchmarks with multi-language, multi-domain task variants is a promising direction for future work.

\subsection{What does concept-level intervention reveal about routing bottlenecks? \label{eval: correction}}
Routesplain’s two-stage design lets us perform a controlled counterfactual: at inference time we replace the concept classifier’s predictions for a single concept group with ground-truth labels, leave every other concept prediction unchanged, and then run the model suitability predictor. Because only one factor is altered, any change in downstream accuracy can be attributed unambiguously to that group. The concept-level prediction performance by group is shown in~\autoref{tab: concept perf}, and the accuracy achieved after each oracle substitution across three cost regularization levels ($\lambda = 0, 0.1, 4.0$) is reported in~\autoref{tab: ablation intervention} (right column).

\textbf{Complexity concepts are the primary bottleneck for routing performance.} Providing oracle complexity labels raises routing accuracy by 3.04--4.05 pp across all cost sensitivities, with the largest gain at $\lambda=0$. Substituting ground-truth labels for library or programming language concepts, however, does not yield any improvement, even though these groups are not predicted perfectly (\autoref{tab: concept perf}). Hence, errors in estimating query complexity propagate directly into routing mistakes, whereas errors in the other concept groups are largely absorbed by the second-stage classifier. Although traditional program analysis metrics (cyclomatic complexity, number of lines of code, Halstead difficulty~\citep{halstead_elements_1979}) do not apply to all software-related tasks, we investigated whether they could replace model-based complexity for code generation tasks but found no improvement. We believe this is because the difficulty of BigCodeBench stems from chaining complex library calls (semantic complexity) rather than syntactic complexity, reflecting real-world code. This underscores the challenge of measuring task complexity even for narrowly defined tasks.

\subsection{Does concept-level intervention lead to predictable routing decisions?}
\label{eval:predictability}
To test whether flipping a single concept produces the intuitive routing shift, we generate 1,000 synthetic concept vectors for each of five source-target programming language pairs (e.g., PHP to Rust). All vectors specify code completion, English, the general domain, and no libraries; query-complexity values are sampled uniformly. The only difference between the paired vectors is that the programming language concept is set to the source language in one case and to the target language in the other. Passing both versions through Routesplain without cost regularization lets us measure how the selection of target-specific models responds to the language flip. Across the five pairs, changing the language concept increases the aggregated selection probability of the top three target-language models for the given language by 37.03 percentage points and improves their average rank by 2.65 positions. For instance, when the programming language concept is changed to PHP, routing shifts toward Llama models, which achieve above 93\% accuracy on PHP (\autoref{fig: intratask perf nl pl}b); when changed to Rust, routing favors o-series models, which show particular strength on that language. The strong, consistent shift confirms that the router’s decisions are directionally consistent in concept space: altering only the programming language concept predictably steers probability mass toward models specialized for that language, validating Routesplain’s controllability through explicit concept manipulation. This experiment also highlights how counterfactual interventions can be used to extract explanations for why specific models are preferred for specific query characteristics.

\section{Conclusion}
We show that routing decisions for software-related tasks can be decomposed into human-interpretable concepts such as task type, programming language, domain, and complexity without sacrificing performance vis-à-vis black-box alternatives. Our proposed Routesplain matches or outperforms all baseline LLM routers and 16 standalone LLMs across eight tasks while providing faithful rationales. Its concept-based design enables diagnostic analyses and interventions unavailable with opaque routers: we identify complexity prediction as the primary routing bottleneck, providing a concrete direction for future improvement.

\section{Limitations}
Routesplain has several practical constraints. The approach requires concept labels for training data and may lack representational power to capture nuanced query relationships not well-described by our concept taxonomy. Additionally, current evaluation datasets often lack diversity across multiple concept dimensions; for example, we leverage an execution-based open-domain Python dataset (BigCodeBench) and an execution-based simple standard library multi-programming-language dataset (MultiPL-E), but no execution-based open-domain multi-programming-language dataset currently exists. This limited cross-concept variability may constrain Routesplain's learning. The system requires supervision and retraining when adding or removing models, adjusting concepts, or updating model pricing. However, retraining costs are negligible compared to LLM inference, as Routesplain trains from scratch in approximately one minute on a single NVIDIA A100.

\section{Reproducibility Statement}
To ensure reproducibility of our results, we provide comprehensive implementation details and experimental specifications. Complete hyperparameter configurations for all router architectures (Routesplain, MLP router, KNN router, and EmbedLLM) are provided in~\autoref{appendix: impl details}, including hidden dimensions, dropout probabilities, learning rates, and batch sizes determined through systematic grid search. Our evaluation uses publicly available datasets (BigCodeBench, CRUXEval, MMLU-ProX/Computer Science, and MultiPL-E) with licensing information detailed in~\autoref{app: licensing information}. For our constructed repair tasks, we provide exact prompt templates in~\autoref{appendix: bcb repair templates} and describe our deterministic filtering methodology for non-trivially incorrect code snippets in~\autoref{subsec: dataset construction}. All experiments follow standard best practices with 80/10/10 train/validation/test splits, 5 independent training runs with reported means and standard deviations, greedy decoding for consistency, and 0-shot (2-shot for CRUXEval) pass@1 accuracy evaluation. We specify the exact embedding models used (codesage-small-v2 and multilingual-e5-small), hardware (single NVIDIA A100 80GB GPU), and statistical testing procedures (two-tailed t-tests and Mann-Whitney U tests). The 38,685 evaluation examples span 16 state-of-the-art LLMs across eight software-related tasks, with model pricing and cost calculations transparently reported in~\autoref{tab: pricing}. Our concept taxonomy and routing strategy are clearly defined in~\autoref{sec: router formulation}, enabling replication of our concept-based routing approach.

\bibliography{references}
\bibliographystyle{colm2026_conference}

\appendix
\section{Implementation Details \label{appendix: impl details}}
\subsection{Routesplain}
Routesplain consists of two separately trained classifiers. Each classifier is a two-layer MLP with GELU~\citep{hendrycks2023gaussianerrorlinearunits} and dropout. We conduct a separate hyperparameter grid search over plausible hidden dimensions, dropout probability, learning rate, and batch size for each classifier and select the combination that results in the lowest mean validation set loss across 3 random seeds. The selected concept-classifier configuration is: hidden dimension = 256, dropout probability = 0.1, learning rate = 0.001, and batch size = 24. For the model classifier, this is hidden dimension = 176, dropout probability = 0.0, learning rate = 0.001, and batch size = 8.

\subsection{MLP Router}
The MLP router is a two-layer MLP classifier with GELU and dropout. We conduct a thorough hyperparameter grid search over plausible hidden dimensions, dropout probability, learning rate, and batch size and select the combination that results in the lowest mean validation set loss across 3 random seeds: hidden dimension = 384, dropout probability = 0.2, learning rate = 0.001, batch size = 8.

\subsection{KNN Router}
For the KNN router, we use scikit-learn's KNeighborsClassifier~\citep{scikit-learn} and select the number of neighbors that results in the highest accuracy on the validation set: $k=20$.

\subsection{EmbedLLM Router}
For the EmbedLLM router, we reimplement the approach as described in~\citet{zhuang2025embedllm} and implemented in the accompanying GitHub repository. We conduct a thorough hyperparameter search over plausible hidden dimensions, model embedding dimensions, learning rate, and batch size and select the combination that results in the lowest validation set loss across 3 random seeds: hidden dimension = 512, model embedding dimension = 128, learning rate = 0.001, batch size = 32.

\section{Comparison of classification accuracy between Routesplain and MLP router \label{app: routesplain mlp acc comp}}
\autoref{tab: routesplain mlp acc comp} summarizes the classification accuracy of Routesplain and the MLP router across varying levels of the cost regularizer $\lambda$. For each setting, we report the mean accuracy and standard deviation over five runs, and provide $p$-values from both a two-tailed $t$-test and a Mann-Whitney U test to assess statistical significance.

Our results demonstrate that the two models achieve comparable accuracy across almost all regularization levels. Statistically significant differences between their means are only observed at $\lambda=0.0$, $0.3$, $0.4$, and $3.0$ ($p<0.05$ for both tests), with the MLP router performing slightly better at low $\lambda$ and Routesplain sometimes outperforming the MLP router at higher $\lambda$. These findings highlight the near parity in accuracy between Routesplain and the MLP router, and importantly, indicate that our interpretable approach can match or surpass the performance of a traditional black-box alternative, especially as cost regularization is increased.

\begin{table*}
    \centering
    \begin{tabular}{ccccc}
        \hline
        $\lambda$ & Routesplain Acc (\%) & MLP Router Acc (\%) & $t$-test $p$ & Mann-Whitney $p$ \\
        \hline
        0.0  & 62.27 {\tiny (0.32)} & 63.46 {\tiny (0.76)} & 0.021 & 0.036 \\
        0.1  & 62.10 {\tiny (0.55)} & 62.71 {\tiny (0.81)} & 0.210 & 0.222 \\
        0.2  & 62.02 {\tiny (0.38)} & 62.40 {\tiny (0.37)} & 0.148 & 0.151 \\
        0.3  & 61.20 {\tiny (0.50)} & 61.94 {\tiny (0.50)} & 0.047 & 0.056 \\
        0.4  & 60.54 {\tiny (0.27)} & 61.46 {\tiny (0.40)} & 0.004 & 0.008 \\
        0.5  & 60.37 {\tiny (0.39)} & 60.74 {\tiny (0.58)} & 0.281 & 0.310 \\
        0.6  & 60.03 {\tiny (0.38)} & 60.07 {\tiny (0.38)} & 0.852 & 0.841 \\
        0.7  & 59.61 {\tiny (0.36)} & 59.78 {\tiny (0.75)} & 0.665 & 0.690 \\
        0.8  & 59.28 {\tiny (0.55)} & 59.28 {\tiny (0.56)} & 0.989 & 1.000 \\
        0.9  & 59.34 {\tiny (0.49)} & 58.90 {\tiny (0.61)} & 0.248 & 0.346 \\
        1.0  & 58.58 {\tiny (0.59)} & 58.61 {\tiny (0.30)} & 0.907 & 0.834 \\
        2.0  & 56.62 {\tiny (0.43)} & 56.20 {\tiny (0.57)} & 0.222 & 0.222 \\
        3.0  & 55.45 {\tiny (0.60)} & 54.32 {\tiny (0.67)} & 0.024 & 0.032 \\
        4.0  & 53.98 {\tiny (0.39)} & 53.70 {\tiny (1.15)} & 0.629 & 1.000 \\
        5.0  & 51.79 {\tiny (0.64)} & 52.53 {\tiny (1.55)} & 0.365 & 0.222 \\
        6.0  & 51.75 {\tiny (0.61)} & 50.83 {\tiny (1.77)} & 0.323 & 0.548 \\
        7.0  & 49.63 {\tiny (1.10)} & 49.98 {\tiny (1.78)} & 0.723 & 0.841 \\
        8.0  & 49.65 {\tiny (0.38)} & 50.11 {\tiny (0.75)} & 0.272 & 0.462 \\
        9.0  & 48.50 {\tiny (0.88)} & 48.90 {\tiny (0.60)} & 0.437 & 0.461 \\
        10.0 & 48.19 {\tiny (1.06)} & 47.64 {\tiny (1.72)} & 0.564 & 0.690 \\
        \hline
    \end{tabular}
    \caption{Comparison of classification accuracy (\%) between Routesplain and MLP router across varying values of the cost regularizer $\lambda$. For each $\lambda$, both models are retrained five times; we report the mean accuracy and standard deviation in parentheses. To assess statistical significance between model performances at each $\lambda$, we report the $p$-values from both a two-tailed $t$-test and the Mann-Whitney U test. This analysis provides insight into the differences in accuracy across different degrees of cost regularization.
}
    \label{tab: routesplain mlp acc comp}
\end{table*}

\section{Concept Prediction Performance}
\autoref{tab: concept perf} reports the prediction quality for each concept group used in Routesplain's concept bottleneck. Binary concept groups (tasks, domains, libraries, natural languages, programming languages) are evaluated using accuracy, precision, recall, and F1. Complexity concepts, which represent continuous ratios, are evaluated using mean squared error (MSE) and mean absolute error (MAE). See~\autoref{eval: correction} for analysis of how concept prediction errors impact routing performance.

\begin{table*}
\centering
\begin{adjustbox}{max width=\textwidth}
\begin{tabular}{c|rrrrrr}
\toprule
    \textbf{Concept Group} & \textbf{Acc (\%)} & \textbf{Prec (\%)} & \textbf{Rec (\%)} & \textbf{F1 (\%)} & \textbf{MSE} & \textbf{MAE} \\ \hline
    Tasks                 &  99.92 \tiny (0.02) & 99.79 \tiny (0.08) & 99.50 \tiny (0.07) & 99.64 \tiny (0.05) &  --- &  --- \\
    Domains               &  99.99 \tiny (0.01) & 99.66 \tiny (0.11) & 99.87 \tiny (0.05) & 99.76 \tiny (0.05) &  --- &  --- \\
    Libraries             &  99.99 \tiny (0.00) & 75.57 \tiny (0.56) & 75.30 \tiny (0.78) & 75.32 \tiny (0.67) &  --- &  --- \\
    Natural Languages     & 100.00 \tiny (0.00) & 99.98 \tiny (0.04) & 99.61 \tiny (0.12) & 99.79 \tiny (0.08) &  --- &  --- \\
    Programming Languages &  99.90 \tiny (0.03) & 99.38 \tiny (0.50) & 91.79 \tiny (2.06) & 94.49 \tiny (1.77) &  --- &  --- \\
    Complexity            &    --- &    --- &   --- &   --- & 0.057 \tiny (0.005) & 0.176 \tiny (0.014) \\
\bottomrule
\end{tabular}
\end{adjustbox}
\caption{\label{tab: concept perf} Concept prediction performance by group. We present accuracy, precision, recall, and F1 for binary labels, mean squared error and mean absolute error for complexity ratios, with mean and standard deviation over 5 runs.}
\end{table*}

\section{Licensing Information \label{app: licensing information}}
We include the licensing information for the datasets and embedding models used in~\autoref{tab:licensing-info}.

\begin{table*}
\centering
\begin{tabular}{l|l|l}
\textbf{Category}   & \textbf{Name}                & \textbf{License}       \\ \hline
Datasets           & BigCodeBench                 & Apache 2.0            \\
                   & CRUXEval                     & MIT                   \\
                   & MMLU-ProX                    & MIT                   \\
                   & MultiPL-E                    & MIT                   \\ \hline
Embedding Models   & intfloat/multilingual-e5-small & MIT                 \\
                   & codesage/codesage-small-v2     & Apache 2.0           \\
\end{tabular}
\caption{Licensing information for the datasets and embedding models used.}
\label{tab:licensing-info}
\end{table*}

\section{BigCodeBench-Repair Prompt Templates \label{appendix: bcb repair templates}}
We include the prompt templates for BigCodeBench-Repair (\autoref{fig:bcb-repair}) and BigCodeBench-Repair+Info (\autoref{fig:bcb-repair+info}).
\begin{figure}
    \begin{tcolorbox}[
        width=\columnwidth,
        colback=white,
        colframe=black!20,
        boxrule=0.5pt,
        arc=1mm]
        \begin{small}
        \texttt{User:}\\
        \> \texttt{I want to solve the following problem:}\\
        \> \texttt{```}\\
        \> \texttt{\{prompt\}}\\
        \> \texttt{```}\\[2pt]
        \\
        \> \texttt{I produced the following code, but it does not work:}\\
        \> \texttt{```python}\\
        \> \texttt{\{code\}}\\
        \> \texttt{```}\\[2pt]
        \\
        \> \texttt{Please provide the fixed code as a self-contained Python script in a Markdown code block.}
        \end{small}
    \end{tcolorbox}
    \caption{Prompt Template for BigCodeBench-Repair.}
    \label{fig:bcb-repair}
\end{figure}
\begin{figure}
    \begin{tcolorbox}[
        width=\columnwidth,
        colback=white,
        colframe=black!20,
        boxrule=0.5pt,
        arc=1mm]
        \begin{small}
        \texttt{User:}\\
        \> \texttt{I want to solve the following problem:}\\
        \> \texttt{```}\\
        \> \texttt{\{prompt\}}\\
        \> \texttt{```}\\[2pt]
        \\
        \> \texttt{I produced the following code, but it does not work:}\\
        \> \texttt{```python}\\
        \> \texttt{\{code\}}\\
        \> \texttt{```}\\[2pt]
        \\
        \> \texttt{Here are more details:}\\
        \> \texttt{```}\\
        \> \texttt{\{additional execution failure information\}}\\
        \> \texttt{```}\\[2pt]
        \\
        \> \texttt{Please provide the fixed code as a self-contained Python script in a Markdown code block.}
        \end{small}
    \end{tcolorbox}
    \caption{Prompt Template for BigCodeBench-Repair+Info.}
    \label{fig:bcb-repair+info}
\end{figure}

\section{LLM Usage}
We used an LLM assistant as a writing aid to fix typos and improve phrasing of human-written text.

\end{document}